\documentclass[journal=jacsat, manuscripts=article]{achemso}

\usepackage{graphicx}
\usepackage{dcolumn}
\usepackage{bm}

\usepackage[utf8]{inputenc}
\usepackage[T1]{fontenc}
\usepackage{mathptmx}
\usepackage{etoolbox}
\usepackage[]{natbib}
\usepackage{array}

\usepackage{mathtools}
\usepackage{autobreak}
\usepackage{bm}
\setkeys{Gin}{width=\linewidth}
\usepackage{wrapfig}
\usepackage{float}
\usepackage{physics}
\usepackage{url}
\usepackage[colorlinks=true,allcolors=black]{hyperref} 
\usepackage{siunitx}
\usepackage{datetime}
\usepackage{enumerate}
\usepackage[version=4]{mhchem}
\usepackage{chemfig}

\renewcommand*\inf{\infty}
\newcommand*\tn{\tilde{\nu}}

\author{Ian F. Mochida}
\email{ian-fitch-mochida@g.ecc.u-tokyo.ac.jp}
\affiliation{Komaba Institute for Science and Department of Basic Science, The University of Tokyo, 3-8-1 Komaba, Meguro, Tokyo 153-8902, Japan}
\author{Tetsuyuki Takayama}%
\author{Shoichi Yamaguchi}
\affiliation{Department of Applied Chemistry, Graduate School of Science and Engineering, Saitama University, 255 Shimo-Okubo,
Sakura, Saitama 338-8570, Japan}
\author{Tetsuya Hama}
\affiliation{Komaba Institute for Science and Department of Basic Science, The University of Tokyo, 3-8-1 Komaba, Meguro, Tokyo 153-8902, Japan}
\email{hamatetsuya@g.ecc.u-tokyo.ac.jp}

\title{An Extended Mixed Quantum/Classical Approach for Quantitative Calculation of Complex Refractive Index}

\begin{document}
\maketitle

\section{AUTHOR INFORMATION}
\textbf{Corresponding Authors:} Ian F. Mochida and Tetsuya Hama

Email: \url{ian-fitch-mochida@g.ecc.u-tokyo.ac.jp} and \url{hamatetsuya@g.ecc.u-tokyo.ac.jp}

Phone: +81-3-5452-6288

\begin{abstract}
    The mixed quantum/classical approach of Skinner and co-workers has been widely used to calculate the line shapes of the infrared spectra of water (\ce{H2O}),
    but less attention has been paid to the use of this approach in quantitatively calculating spectral intensity, thereby limiting direct comparisons of calculated and experimental spectra.
    Here, we extend this theoretical framework to facilitate direct computation of the full complex refractive index of water, replacing the normalized ordinate used in previous studies.
    Our results for the OH stretching region of \ce{H2O} capture both the shapes and intensities of the experimental spectra.
    They reveal that inclusion of the local field effect is crucial to the accurate reproduction of spectral intensity.
    This extended approach enables new areas of analysis of the bulk, thin-film, and cluster spectra of water.
\end{abstract}

Vibrational spectra of condensed-phase water directly reflect the environment of each molecule and provide rich information on the structure\cite{falkINFRAREDSPECTRUMSTRUCTURE1966,klugHighdensityAmorphousIce1987,brubachSignaturesHydrogenBonding2005,bakkerVibrationalSpectroscopyProbe2010,perakisVibrationalSpectroscopyDynamics2016,sekiBendingModeWater2020}, inter- and intra-molecular couplings \cite{sokolowskaCrossingAnisotropicIsotropic1993,woutersenResonantIntermolecularTransfer1999,toriiUltrafastAnisotropyDecay2000,auerIRRamanSpectra2008,yangSignaturesCoherentVibrational2010,mattInfluenceIntermolecularCoupling2018,yuVibrationalCouplingsEnergy2020,perakisVibrationalSpectroscopyDynamics2016,ramaseshaWaterVibrationsHave2013,demarcoAnharmonicExcitonDynamics2016,lindnerVibrationalRelaxationPure2006}, and dynamics\cite{carpenterDelocalizationStretchbendMixing2017,perakisVibrationalSpectroscopyDynamics2016,bakkerVibrationalSpectroscopyProbe2010,nihonyanagiUltrafastDynamicsWater2017}.
To interpret experimental spectra, many approaches have been developed to calculate the theoretical vibrational spectra of water \cite{guillotMolecularDynamicsStudy1991,toriiInfluenceLiquidDynamics2002,toriiTimeDomainCalculationsPolarized2006,jansenDissimilarDynamicsCoupled2009,choiComputationalIRSpectroscopy2013,burnhamVibrationalProtonPotential2008,ishiyamaAnalysisAnisotropicLocal2009,hasegawaPolarizableWaterModel2011,wangSystematicImprovementClassical2013,meddersInfraredRamanSpectroscopy2015,spuraNuclearQuantumEffects2015,liInitioVibrationalSpectroscopy2025}.

One of which is the so-called mixed quantum/classical approach, which encompasses theoretical approaches that treat low-frequency translational and rotational modes classically and high-frequency bending and stretching modes quantum mechanically, in which way effectively lowers the computational cost.
For example, Torii introduced the wave function propagation (WFP) method,\cite{toriiInfluenceLiquidDynamics2002,toriiTimeDomainCalculationsPolarized2006} and Jansen and Knoester reported the scheme for numerical integration of the Schrödinger equation (NISE)\cite{jansenNonadiabaticEffectsTwoDimensional2006,jansenDissimilarDynamicsCoupled2009}.
Although these approaches can calculate vibrational spectra by accurately incorporating dynamical effects, they still require relatively high computational costs that often make them impractical for routine use.
Therefore, a more efficient approach that maintains reasonable accuracy is desirable.

To this end, Skinner and co-workers have developed their version of the mixed quantum/classical approach that compellingly balances accuracy and efficiency\cite{auerDynamicalEffectsLine2007,auerIRRamanSpectra2008,kananenkaFermiResonanceOHstretch2018}.
The matrix elements of the vibrational Hamiltonian and local mode transition dipole moments are provided by vibrational spectroscopic maps\cite{gruenbaumRobustnessFrequencyTransition2013,kananenkaFermiResonanceOHstretch2018,baizVibrationalSpectroscopicMap2020}.
These maps are linear or quadratic functions fitted to a large set of results of electronic structure calculations of small clusters taken from molecular dynamics (MD) simulations.
Although their initial preparation is computationally demanding, these maps greatly reduce the cost of further calculations by avoiding the need for repeated quantum mechanical calculations for every configuration.
As reviewed by Baiz et al.\cite{baizVibrationalSpectroscopicMap2020}, this substantial improvement in efficiency led to extensive exploration of the capability of this tool.
For example, Skinner and co-workers developed vibrational spectroscopic maps of water, which they used to successfully interpret the vibrational spectra of water\cite{auerIRRamanSpectra2008,kananenkaFermiResonanceOHstretch2018}.
Before solving the time-independent Schrödinger equation, the vibrational Hamiltonian is often averaged over time $T$, a technique termed the time-averaging approximation (TAA)\cite{auerDynamicalEffectsLine2007,auerIRRamanSpectra2008,yangTimeaveragingApproximationInteraction2011,takayamaExperimentalTheoreticalRaman2022,takayamaTransferabilityVibrationalSpectroscopic2023,yamaguchiAppraisalTIP4PtypeModels2023,takayamaTheoreticalExperimentalODstretch2024,takayamaSimulationStudyRaman2025}.
This approximation can substantially reduce computational cost\cite{yangTimeaveragingApproximationInteraction2011} while incorporating dynamical effects to maintain an accuracy comparable to that of more rigorous time-domain approaches such as WFP\cite{toriiInfluenceLiquidDynamics2002,toriiTimeDomainCalculationsPolarized2006} and NISE\cite{jansenNonadiabaticEffectsTwoDimensional2006,jansenDissimilarDynamicsCoupled2009}.
By combining vibrational spectroscopic maps, TAA, and classical MD simulations of the cost-effective, rigid, non-polarizable water model\cite{berendsenMissingTermEffective1987,jorgensenComparisonSimplePotential1983tip4p,abascalGeneralPurposeModel2005,tainterReparametrizedE3BExplicit2015}, Skinner's approach preserves adequate accuracy while substantially lowering the computational cost.
This approach has been used to calculate OH stretching and HOH bending infrared (IR), 2D-IR, Raman, and sum-frequency generation spectra for liquid water, ice, water clusters, electrolyte solutions, and various water isotopologues\cite{corcelliInfraredRamanLine2005,schmidtPronouncedNonCondonEffects2005,schmidtAreWaterSimulation2007,auerIRRamanSpectra2008,liInfraredRamanLine2010,liInfraredRamanLine2010a,yangSignaturesCoherentVibrational2010,jansenTwodimensionalInfraredSpectroscopy2010,yangObtainingInformationProtein2015,shiInterpretationIRRaman2012,tainterHydrogenBondingOHStretch2013,tainterStructureOHstretchSpectroscopy2014,niIRSFGVibrational2015,kananenkaFermiResonanceOHstretch2018,ishiharaRamanSpectroscopyIsotopically2022,takayamaExperimentalTheoreticalRaman2022,takayamaTheoreticalExperimentalODstretch2024,takayamaSimulationStudyRaman2025,yamaguchiApplicationsTheoreticalMethods2025}.

In addition to the line shape, intensity is another important characteristic of vibrational spectra.
For IR absorption spectroscopy in particular, the intensity corresponds to the rate of light absorption and is often expressed through spectroscopic parameters such as absorbance ($A$), absorption cross-section ($\sigma$), and the imaginary part ($k$) of the complex refractive index ($\hat{n} = n + ik$), also known as the extinction coefficient.
The Beer--Lambert law describes the relationship between these quantities at a given angular frequency ($\omega$):\cite{hasegawaQuantitativeInfraredSpectroscopy2017}
\begin{align}
    A(\omega) = -\log_{10} \frac{I(\omega)}{I_0(\omega)} = \frac{N \sigma(\omega)}{\ln 10} = \frac{2 d \omega  k(\omega)}{c \ln 10}, \label{eq:Lambert-Beer}
\end{align}
where $I_0$ and $I$ are the incident and transmitted light intensities, respectively; $N$ is the number density of the absorbing species; $d$ is the path length; and $c$ is the speed of light.

Quantitative calculations of the spectral intensity would facilitate comprehensive comparison of theoretical and experimental results.
In a pioneering work, Guillot calculated the infrared absorption intensity of the spectral bands of liquid water over the range of \qtyrange{0.5}{1000}{\per\cm} using a rigid, non-polarizable water model\cite{guillotMolecularDynamicsStudy1991}.
This region reflects the intermolecular motion, which can be adequately treated by classical dynamics in the theoretical procedure.
The approach has been adopted by others to investigate the intermolecular motion of water using different rigid and non-polarizable water models\cite{bosmaSimulationIntermolecularVibrational1993,zasetskyStudyTemperatureEffect2007,shiDielectricConstantLowfrequency2014,segaDielectricTerahertzSpectroscopy2015}.
However, we know of no work that has extended this efficient approach to the OH stretching region of water, presumably because it would require much computationally demanding quantum mechanical treatment of the faster intramolecular dynamics.
Instead, quantitative IR spectra of water covering this region have been calculated by other methods such as \textit{ab-initio} MD, path-integral MD, or classical MD with polarizable models\cite{burnhamVibrationalProtonPotential2008,ishiyamaAnalysisAnisotropicLocal2009,hasegawaPolarizableWaterModel2011,wangSystematicImprovementClassical2013,meddersInfraredRamanSpectroscopy2015,spuraNuclearQuantumEffects2015,liInitioVibrationalSpectroscopy2025}.
Although the theoretical spectra of these studies present overall reasonable agreement with experimental spectra, their computational cost limits their application to realistic condensed phase systems.
Therefore, it is desirable to extend Skinner's approach to enable the efficient calculation of spectral intensity while preserving its established accuracy for predicting line shapes.

In addition, we know of no direct calculation using this method of the real part of the refractive index ($n$) in the IR range.
Typically, the value of $n(\omega)$ at a given $\omega$ is calculated from experimental or theoretical values of $k(\omega)$ via the Kramers--Kronig (KK) relation\cite{robinsonDeterminationInfraRedAbsorption1953,bertieInfraredIntensitiesLiquids1985,hasegawaQuantitativeInfraredSpectroscopy2017}:
\begin{align}
    k(\omega)          & = \frac{-2\omega}{\pi} P\int_0^\inf \dd{\omega'} \frac{n(\omega')}{\omega'^2 - \omega^2} , \label{eq:KrKk} \\
    n(\omega) - n_\inf & = \frac{2}{\pi} P\int_0^\inf \dd{\omega'}\frac{\omega' k(\omega')}{\omega'^2 - \omega^2}, \label{eq:KKn}
\end{align}
where $P$ indicates the Cauchy principal value and $n_\inf$ is a constant representing the refractive index for near-IR frequencies.
Values of $n(\omega)$ become essential when calculating IR spectra of thin films of various phases of water, including vapor-deposited amorphous water.
The line shape of thin-film spectrum is governed by the transverse optic (TO) energy-loss function ($f(\epsilon)_\text{TO}$) and longitudinal optic (LO) energy-loss function ($f(\epsilon)_\text{LO}$), which are approximately expressed as:\cite{hasegawaQuantitativeInfraredSpectroscopy2017,nagasawaInfraredMultipleangleIncidence2022}
\begin{align}
    f(\epsilon_{xy})_\text{TO} & = \text{Im}\; \epsilon_{xy} = 2n_{xy}k_{xy},                                       \\
    f(\epsilon_z)_\text{LO}    & = \text{Im} \; \qty[-\frac{1}{\epsilon_z}] = \frac{2n_zk_z}{({n_z}^2 +{k_z}^2)^2},
\end{align}
where $\epsilon_{xy}$, $n_{xy}$, and $k_{xy}$ are the surface-parallel components of the dielectric constant, real and imaginary parts of the complex refractive index, respectively, of the thin sample.
$\epsilon_z$, $n_z$, and $k_z$ are the surface-perpendicular components of the corresponding values.
As demonstrated for amorphous water at approximately \SI{10}{K} in Figure~S1, line shapes of normal-incidence transmission and reflection-absorption measurements are governed by $f(\epsilon)_\text{TO}$ and $f(\epsilon)_\text{LO}$, respectively.
The difference between the two functions is due to the strong dispersion of $n(\omega)$ in the vicinity of a strong absorption band causing a peak shift known as the TO-LO splitting.
Therefore, calculating theoretical values of $n(\omega)$ is necessary to directly simulate such thin-film spectra to investigate structural properties of thin-film materials.
Although $n(\omega)$ can in theory be derived from $k(\omega)$ through the KK relation, a substantial challenge facing this approach is the infinite integration range, the truncation of which introduces error in practice.
Furthermore, Skinner's method is limited to spectral regions with available vibrational maps, which increases the potential errors.
Given these limitations, it is desirable to derive equations to calculate $n(\omega)$ directly, thereby circumventing the need for the KK relation when extending Skinner's method.

This study revisits the formulation of Skinner's mixed quantum/classical approach and extends it using linear response theory to introduce a direct expression for the full complex dielectric constant, from which quantitative expressions for both $k(\omega)$ and $n(\omega)$ can be readily obtained.
Our derivation introduces local-field correction (LFC), which relates the optical response of a single molecule to the bulk optical property\cite{lorentzTheoryElectronsIts1909,aubretUnderstandingLocalFieldCorrection2019}.
Although LFC has not been included in previous calculations, we hypothesize that local field effects cause large deviations in calculated intensities from experimental results.
We apply our new formulations to calculate $k(\omega)$ and $n(\omega)$ for liquid water.
Features of the spectra are qualitatively analyzed by taking the second derivative.
Quantitative analysis is also conducted by considering other spectroscopic parameters such as the absorption cross-section, molar absorption coefficient, and band strength.
Subsequently, we explicitly evaluate the effects of LFC on the line shape and intensity by comparing spectra calculated with and without LFC.

Previous studies have expressed the line shape $I(\omega)$ of IR spectra as follows:\cite{auerDynamicalEffectsLine2007,auerIRRamanSpectra2008,kananenkaFermiResonanceOHstretch2018,mcquarrieStatisticalMechanics2000}
\begin{align}
    I(\omega) \propto \text{Re} \int_0^\inf \dd{t} e^{-i\omega t} \ev**{\hat{\epsilon} \cdot \hat{\mu}(0) \hat{\mu}(t) \cdot \hat{\epsilon}}, \label{eq:prev}
\end{align}
where $\hat{\epsilon}$ is the laboratory-fixed direction of polarization of the incident light, $\hat{\mu}(t)$ is the Heisenberg expression of the total dipole operator of the system at time $t$, and angular brackets denote a quantum equilibrium statistical mechanical average.
This formulation is expressed in a normalized form, representing no specific physical quantity.

We now quantitatively extend eq~\eqref{eq:prev} using linear response theory to obtain quantitative expressions of $k(\omega)$ and $n(\omega)$.
Consider $N$ coupled two-state chromophores (OH bonds in the current case) within a volume $V$.
In linear vibrational spectroscopy at room temperature, the transition energy $\hbar \omega$ is sufficiently larger than the thermal energy ($k_\text{B} T \ll \hbar \omega$) which allows a reasonable assumption of most vibrational states to be in their ground state.
Irradiating the system with a beam of light with an electric field $\vec{E}(\omega)$ results in the polarization of the system $\vec{P}(\omega)$, given by
\begin{align}
    \vec{P}(\omega) = \epsilon_0 (\epsilon_r(\omega) - 1) \vec{E}(\omega), \label{eq:def-chi}
\end{align}
where $\epsilon_0$ is the vacuum dielectric constant and $\epsilon_r(\omega)$ is the wavenumber-dependent dielectric constant of the system.
Based on linear response theory\cite{guillotMolecularDynamicsStudy1991,mcquarrieStatisticalMechanics2000,yamaguchiApplicationsTheoreticalMethods2025}, $\epsilon_r(\omega)$ is expressed in the vibrational region as
\begin{align}
    \epsilon_r(\omega) = \epsilon_b + \frac{1}{3} \sum_{p=x,y,z}\frac{i}{\epsilon_0 \hbar V}\int_0^{\inf} \dd{t} e^{-i\omega t} \ev**{\hat{\epsilon}_p \cdot \hat{\mu}(0)\hat{\mu}(t)\cdot \hat{\epsilon}_p}, \label{eq:chi}
\end{align}
where $\epsilon_b$ is the background dielectric constant, which originates from purely electronic polarization.
$\hat{\epsilon}_p$ are the unit vector in the direction of $p$, which is one of the three Cartesian coordinates ($x, y, z$).
An average over the three directions is taken for the isotropic system to improve the signal-to-noise ratio.
The angular brackets denote a quantum equilibrium statistical mechanical average.
Eq~\eqref{eq:prev} is now quantitatively expressed in the second term of eq~\eqref{eq:chi}.
The equation can be further transformed using previous results from Skinner's group.
Within the mixed quantum/classical approach, $\epsilon_r(\omega)$ can be expressed as\cite{auerDynamicalEffectsLine2007}
\begin{align}
    \epsilon_r(\omega) = \epsilon_b + \frac{1}{3} \sum_{p=x,y,z} \frac{i}{\epsilon_0 \hbar V} \int_0^\inf \dd{t} e^{-i\omega t} \sum_{i, j} \ev**{m_{ip}(0)F_{ij}(t)m_{jp}(t)} e^{-t/2T_1}, \label{eq:eps-1}
\end{align}
where $m_{ip}(t)$ are the inner products of the transition dipoles of chromophores $i$ ($\hat{\mu}_i$) and $\hat{\epsilon}_p$, expressed as $m_{i}(t) = \hat{\mu}_{ip}(t) \cdot \hat{\epsilon}_p$, and $F_{ij}(t)$ are the elements of the matrix $F(t)$, which satisfy the equations
\begin{align}
    \dot{F}(t) = iF(t)\kappa(t), \quad  F_{ij}(0) = \delta_{ij}, \label{eq:F}
\end{align}
where $\kappa(t)$ is a matrix whose elements are expressed as
\begin{align}
    \kappa_{ij}(t) = \omega_i(t)\delta_{ij} + \omega_{ij}(t)(1-\delta_{ij}).
\end{align}
Here, $\omega_i(t)$ are the fluctuating transition frequencies, and $\omega_{ij}(t)$ are the fluctuating couplings.
The angular brackets now denote a classical equilibrium average.
$T_1$ is a phenomenological parameter that reflects the lifetime effects of the OH stretching excitation.
Under TAA\cite{auerDynamicalEffectsLine2007}, $\kappa(t)$ is averaged over time $T$ as
\begin{align}
    \kappa_T = \frac{1}{T} \int_0^T \dd{t} \kappa(t),
\end{align}
and $m_{ip}(t)$ is replaced with $m_{ip}(T)$.
Setting $T$ to \SI{76}{fs} has been demonstrated to best reproduce the line shape of liquid water\cite{auerDynamicalEffectsLine2007}, which will be the case in the present study.
$\kappa_T$ is diagonalized by the orthogonal transformation $M^T \kappa_T M$ with a square matrix $M$.
Lastly, we define $d_{kp}(t)$ as
\begin{align}
    d_{kp}(t) = \sum_i m_{ip} (t) M_{ik} \label{eq:d_k}
\end{align}
and $\gamma_k$ as the eigenvalues of $\kappa_T$.
Replacing $m_{ip}(t)$ in eq~\eqref{eq:eps-1} with $m_{ip}(T)$ and then substituting eq~\eqref{eq:d_k} into eq~\eqref{eq:eps-1} gives the following formulation of $\epsilon_r(\omega)$:
\begin{align}
    \epsilon_r(\omega) = \epsilon_b + \frac{1}{3} \sum_{p=x,y,z}\frac{1}{\epsilon_0 \hbar V} \sum_{k} \ev**{d_{kp}(0) d_{kp}(T) L(\omega - \gamma_k)}. \label{eq:eps-pre-loc}
\end{align}
Here, $L(\omega)$ is the complex Lorentzian function, expressed as
\begin{align}
    L(\omega) & = \frac{1}{\omega - i/(2T_1)}                                                                           \\
              & = \frac{\omega}{\omega^2 + (1/(2T_1))^2} + \frac{i/(2T_1)}{\omega^2 + (1/(2T_1))^2}. \label{eq:lorentz}
\end{align}
By combining linear response theory and Skinner's previous formulation, we have arrived at a new complex and quantitative expression (eq~\eqref{eq:eps-pre-loc}) that replaces the previous line shape function (eq~\eqref{eq:prev}), which fully describes the optical properties of the system.

We next consider the effects of the local field.
A bulk medium in an external electric field $\vec{E}_\text{ext}(\omega)$ will have its constituent molecules experience a different, local electric field $\vec{E}_\text{loc}(\omega)$.
The LFC is an adjustment that accounts for this local field effect \cite{lorentzTheoryElectronsIts1909,aubretUnderstandingLocalFieldCorrection2019}.
Theoretical calculations of the quantitative IR spectra of water usually utilize polarizable water models\cite{burnhamVibrationalProtonPotential2008,ishiyamaAnalysisAnisotropicLocal2009,hasegawaPolarizableWaterModel2011,wangSystematicImprovementClassical2013,meddersInfraredRamanSpectroscopy2015,spuraNuclearQuantumEffects2015,liInitioVibrationalSpectroscopy2025}, thus implicitly including LFC\cite{moritaMolecularTheoryLocal2018}.
However, the computationally efficient non-polarizable models employed here do not incorporate polarization effects, so LFC must be explicitly applied to the formulation.

In eq~\eqref{eq:eps-1}, $\epsilon_r(\omega)$ represents the system's response to $\vec{E}_\text{loc}(\omega)$, whereas the experimental dielectric constant $\epsilon_\text{cor}(\omega)$ is evaluated with respect to $\vec{E}_\text{ext}(\omega)$.
Therefore, $\epsilon_r(\omega)$ in eq~\eqref{eq:eps-pre-loc} must be adjusted to account for this discrepancy in order to describe accurately the experimental dielectric constant.
For a homogeneous and isotropic medium such as pure water, $\vec{E}_\text{loc}(\omega)$ and $\vec{E}_\text{ext}(\omega)$ are related by\cite{aubretUnderstandingLocalFieldCorrection2019}
\begin{align}
    \vec{E}_\text{loc}(\omega) & = \vec{E}_\text{ext}(\omega) + \frac{1}{3\epsilon_0} \vec{P}(\omega)   \\
                               & = \frac{\epsilon_\text{cor}(\omega) + 2}{3}\vec{E}_\text{ext}(\omega).
    \label{eq:loc-ext}
\end{align}
Here, $(\epsilon_\text{cor}(\omega) + 2)/3$ is the Lorentz local field correction factor.
Pairs of $\epsilon_r, \vec{E}_\text{loc}$ and $\epsilon_\text{cor}, \vec{E}_\text{ext}$ can both describe the system's polarization:
\begin{align}
    \vec{P}(\omega) & = \epsilon_0 (\epsilon_r(\omega) - 1) \vec{E}_\text{loc}(\omega),  \label{eq:loc} \\
    \vec{P}(\omega) & = \epsilon_0 (\epsilon_\text{cor}(\omega)  - 1) \vec{E}_\text{ext}(\omega).
    \label{eq:ext}
\end{align}
Substituting eq~\eqref{eq:loc-ext} into eq~\eqref{eq:loc} obtains another relation between $\vec{P}(\omega)$ and $\vec{E}_\text{ext}(\omega)$:
\begin{align}
    \vec{P}(\omega) = \epsilon_0 (\epsilon_r(\omega) -1) \qty(\frac{\epsilon_\text{cor}(\omega) + 2}{3})\vec{E}_\text{ext}(\omega).
    \label{eq:ext-2}
\end{align}
Comparing the coefficients of eqs~\eqref{eq:ext} and \eqref{eq:ext-2} finally yields the expression
\begin{align}
    \epsilon_\text{cor}(\omega) = \frac{2\epsilon_r(\omega) + 1}{4 - \epsilon_r(\omega)}.\label{eq:fin}
\end{align}
We can use eq~\eqref{eq:fin} to apply LFC to the dielectric constant $\epsilon_r(\omega)$ in eq~\eqref{eq:eps-pre-loc} to yield the corrected dielectric constant $\epsilon_\text{cor}(\omega)$.

\begin{figure}
    \centering
    \includegraphics[width=0.8\linewidth]{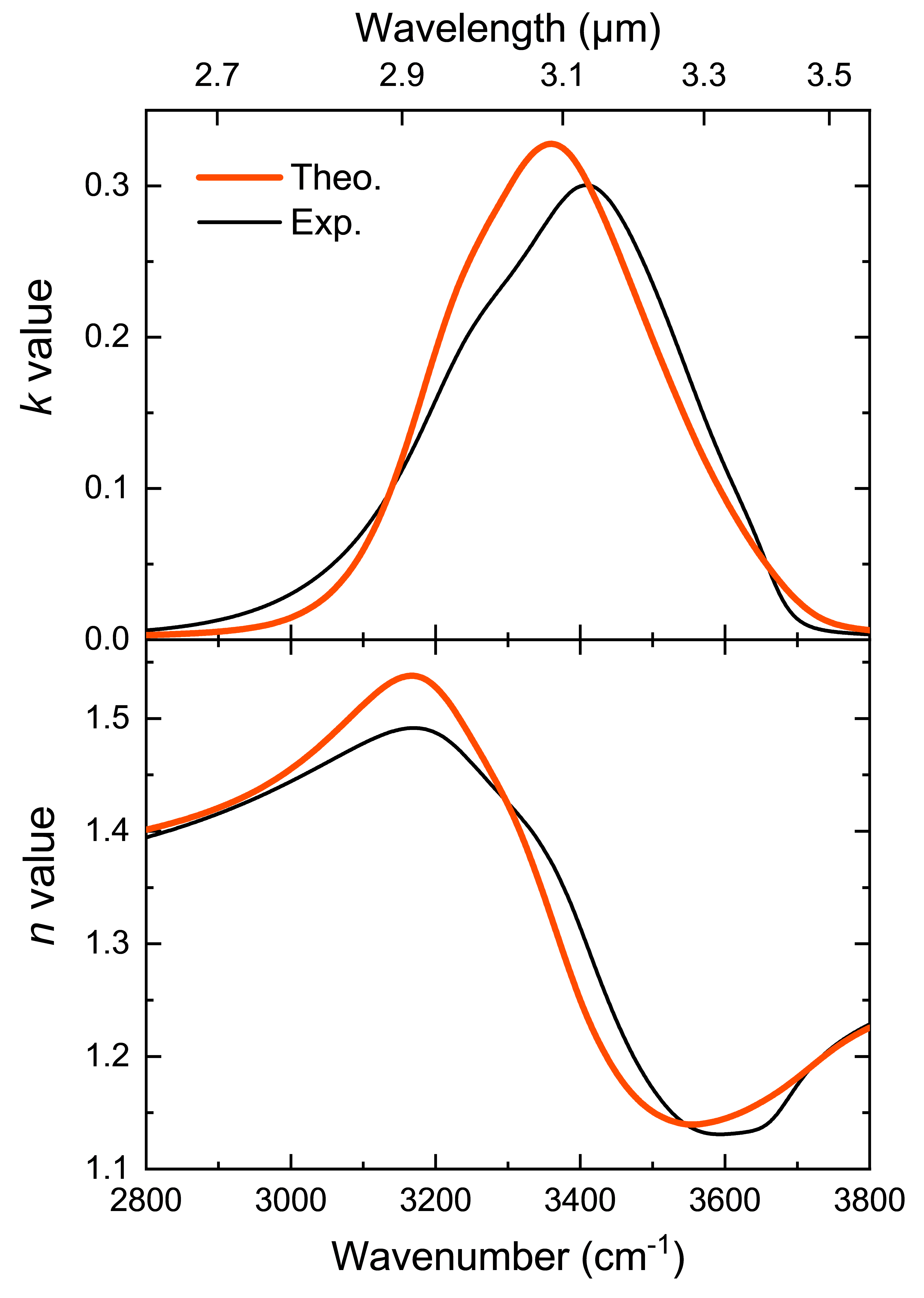}
    \caption{Theoretical (red) and experimental (black) OH stretching spectra of \ce{H2O}. The upper and lower plots respectively show the imaginary ($k$) and real ($n$) parts of the complex refractive index.}
    \label{fig:n-k}
\end{figure}
\begin{figure}
    \centering
    \includegraphics[]{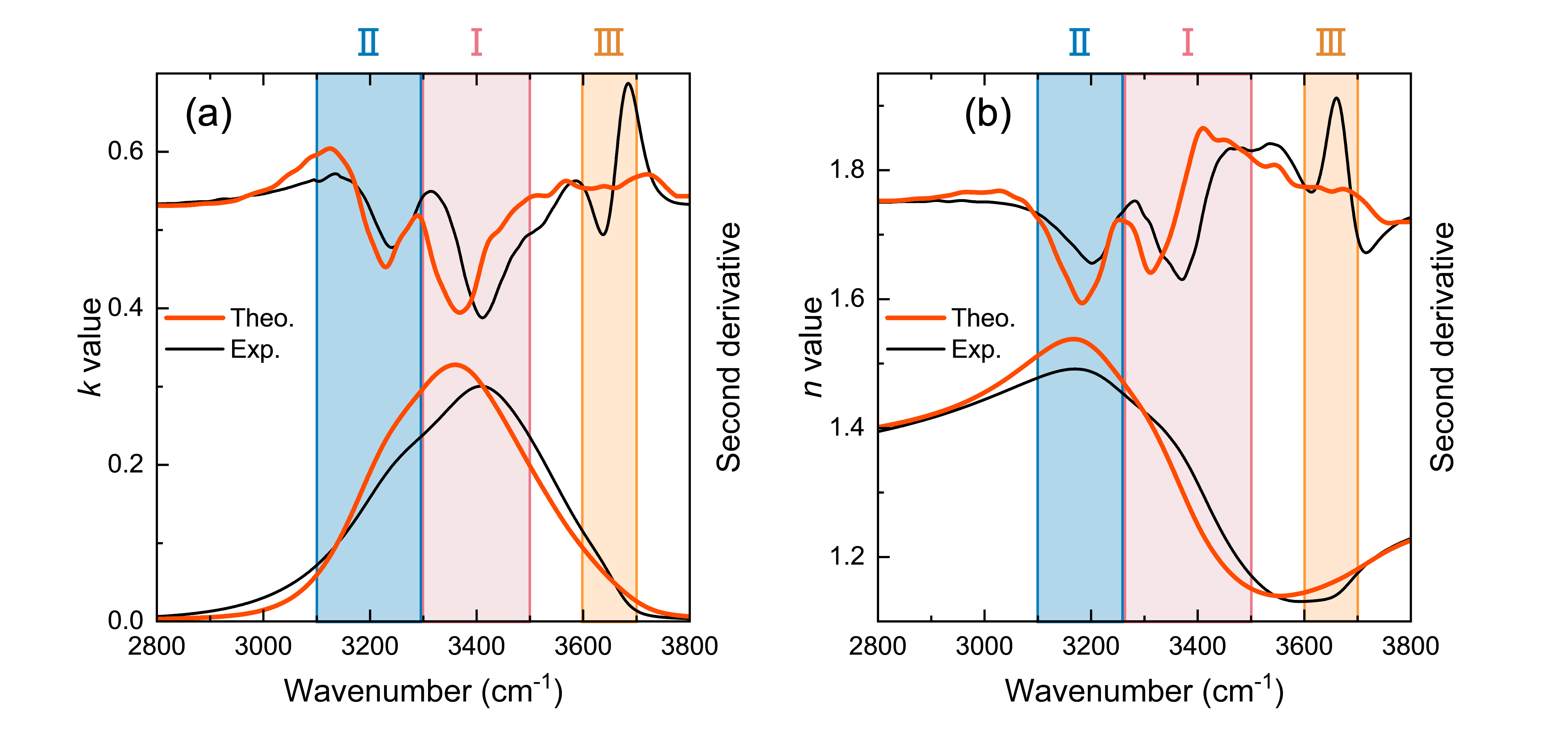}
    \caption{Second derivatives of the theoretical (red) and experimental (black) spectra in Figure~\ref{fig:n-k}, showing the (a) imaginary ($k$) and (b) real ($n$) parts of the complex refractive index.Pink, blue, and orange areas represent regions I, II, and III, respectively. The original spectra are included in the lower part of each plot.}
    \label{fig:n-k-second}
\end{figure}
$k(\omega)$ and $n(\omega)$ can be calculated from the values of $\epsilon_\text{cor}(\omega)$ obtained from eq~\eqref{eq:fin}.
The relation between $\epsilon_\text{cor}(\omega)$ and the complex refractive index $\hat{n}(\omega)$ (where $\hat{n}(\omega)=n(\omega) + ik(\omega))$ is given by
\begin{align}
    \epsilon_\text{cor}(\omega) & = \hat{n}(\omega)^2 = n(\omega)^2-k(\omega)^2+2in(\omega)k(\omega).
\end{align}
From this, we obtain the real and imaginary parts of the dielectric constant ($\epsilon_\text{cor}(\omega) = \epsilon_1(\omega) + i\epsilon_2(\omega)$):
\begin{align}
    \epsilon_1(\omega)=n(\omega)^2-k(\omega)^2,\quad \epsilon_2(\omega)=2n(\omega)k(\omega),
\end{align}
which can be inverted to yield
\begin{align}
    k(\omega) & = \sqrt{\frac{-\epsilon_1(\omega)+\sqrt{\epsilon_1(\omega)^2+\epsilon_2(\omega)^2}}{2}},  \label{eq:k} \\
    n(\omega) & = \sqrt{\frac{\epsilon_1(\omega)+\sqrt{\epsilon_1(\omega)^2+\epsilon_2(\omega)^2}}{2}}, \label{eq:n}
\end{align}
directly providing the components of $\hat{n}(\omega)$.

We note that our formulation provides a more accurate description of the absorption line shape as well.
The imaginary part of $\epsilon(\omega)$ in eq~\eqref{eq:chi}, which equals to $2n(\omega)k(\omega)$, is linearly related to eq~\eqref{eq:prev}.
On the other hand, the absorption spectrum of a bulk medium is proportional solely to $k(\omega)$ (eq~\eqref{eq:Lambert-Beer}).
Thus, direct comparison of eq~\eqref{eq:prev} with experimental absorption spectra, as performed in previous studies\cite{auerIRRamanSpectra2008,liInfraredRamanLine2010,shiInterpretationIRRaman2012,kananenkaFermiResonanceOHstretch2018}, implicitly assumes that $n(\omega)$ is constant across the band.
This assumption is invalid due to the large dispersion of $n(\omega)$ in the vicinity of an absorption band.
Our formulation avoids this assumption by explicitly considering wavenumber-dependent values of $k(\omega)$ and $n(\omega$), thereby achieving a more accurate treatment of the absorption line shape.

In summary, three steps are introduced to calculate the complex refractive index $\hat{n}(\omega)$: (1) calculate $\epsilon_r(\omega)$ using eq~\eqref{eq:eps-pre-loc}, (2) apply LFC to $\epsilon_r(\omega)$ using eq~\eqref{eq:fin} to obtain $\epsilon_\text{cor}(\omega)$, and (3) calculate $k(\omega)$ and $n(\omega)$ from $\epsilon_\text{cor}(\omega)$ using eqs~\eqref{eq:k} and \eqref{eq:n}.
In practice, step (1) is the most computationally expensive, whereas steps (2) and (3) are simple substitutions.

We validate our extended approach by calculating the OH stretching vibrational spectra of liquid water at \SI{298}{K}.
First, we ran a MD simulation using 512 molecules of the TIP4P/2005 water model\cite{abascalGeneralPurposeModel2005}.
For details, see the Computational Details section.
Subsequently, spectra of $k$ and $n$ were calculated using the new formulations after constructing the vibrational Hamiltonian based on prior vibrational spectroscopic maps\cite{gruenbaumRobustnessFrequencyTransition2013,kananenkaFermiResonanceOHstretch2018}.
Figure~\ref{fig:n-k} shows in red the theoretical spectra of $k$ and $n$ for liquid water; the black lines show prior experimental data measured at \SI{300}{K}\cite{maxIsotopeEffectsLiquid2009}.
The theoretical spectra for $k$ and $n$ agree well overall with the experimental data in terms of both intensity and line shape.
In detail, the calculated peak of $k$ is slightly red-shifted and more intense than the experimental peak, a deviation also observed for $n$.
The similarity of these deviations is an expected consequence of the correlation of $k$ and $n$ through the KK relation.
Although our method directly calculates $k$ and $n$ without explicitly invoking this relation, the two quantities are still inherently linked by it, which results in their similar behavior.
The KK relation is rigorously defined by an integral over an infinite range, but its influence is predominantly local, as values of $k$ and $n$ at a given wavenumber are most strongly influenced by each other in the same spectral region due to the weighted integrals in eqs~\eqref{eq:KrKk} and \eqref{eq:KKn}.
Therefore, the KK relation can be effectively applied over a finite range to aid interpretation of the spectra.
The accuracies of the theoretical spectra of $\epsilon_1$ and $\epsilon_2$ resemble those of $n$ and $k$, respectively, because $n$ and $k$ each have dominant contributions to the line shapes of $\epsilon_1$ and $\epsilon_2$ (see Figure~S2).

The line shapes in Figure~\ref{fig:n-k} are further analyzed by taking their second derivatives.
Second-derivative analysis is useful for separating the peak positions of overlapping features\cite{yangObtainingInformationProtein2015,mitsuotasumiIntroductionExperimentalInfrared2015}.
Overlapping features in the original spectrum appear as distinct downward-pointing peaks in the second-derivative spectrum.
Although the OH stretching band of water is known to be better described as a continuum of coupled OH chromophores rather than a few sub-bands\cite{auerIRRamanSpectra2008,yangSignaturesCoherentVibrational2010}, second-derivative analysis can extract subtle features, enabling clear comparison of the spectral features.
Figure~\ref{fig:n-k-second} shows the second derivatives of the theoretical and experimental spectra, displaying them above the original spectra.
The second derivative of $k$ was calculated from normalized spectra with the same peak heights to aid visibility.
The shoulder-like features (around \SI{3230}{\per\cm} for $k$ and \SI{3200}{\per\cm} for $n$) are clearly separated from the main features in the second-derivative spectra, demonstrating the utility of the technique.
Notable features in the second-derivative spectra are colored and labeled I, II, and III.
As peaks in the second-derivative plots for $k$ and $n$ at similar wavenumber ranges are correlated via the KK relation, they share the same labels.

The results for $k$ in Figure~\ref{fig:n-k-second}(a) show that theoretical peak I, corresponding to the main peak position in the original spectrum, is red-shifted by \SI{40}{\per\cm} compared with the experimental peak.
Peak II corresponds to the shoulder-like feature in the original spectrum and is red-shifted by \SI{10}{\per\cm}, a smaller deviation than that of peak I.
In contrast, peak III, which appears as a subtle shoulder in the original experimental spectrum, does not appear in the theoretical spectrum.
This is due to inaccuracies in the vibrational spectroscopic maps\cite{kananenkaFermiResonanceOHstretch2018,takayamaTheoreticalExperimentalODstretch2024}.
The origin of peak III can be regarded as an accidental consequence of the interplay of the structural and dynamical properties of the hydrogen-bond network.
Although OH chromophores with weak or no hydrogen bonds are thought to contribute to the blue side of the OH stretching band\cite{auerHydrogenBondingRaman2007,auerWaterHydrogenBonding2009}, it is generally considered misleading to attribute peak III to a specific molecular environment or vibrational mode\cite{maxIsotopeEffectsLiquid2010,yangSignaturesCoherentVibrational2010}.

For $n$ in Figure~\ref{fig:n-k-second}(b), the deviation from the experimental result is similar to that for $k$.
In the second-derivative spectrum, theoretical peak I, which corresponds to the shoulder-like feature in the original spectrum, is red-shifted by \SI{60}{\per\cm} compared with the experimental results.
Peak II, corresponding to the prominent upward feature in the original spectrum, is red-shifted by \SI{20}{\per\cm}, which is less than that for peak I.
Experimental peak III is a sharp upward peak due to the downward feature in the original spectrum, and its absence from the theoretical spectrum is likely due to the same reason as discussed for $k$.

\begin{figure}
    \centering
    \includegraphics[]{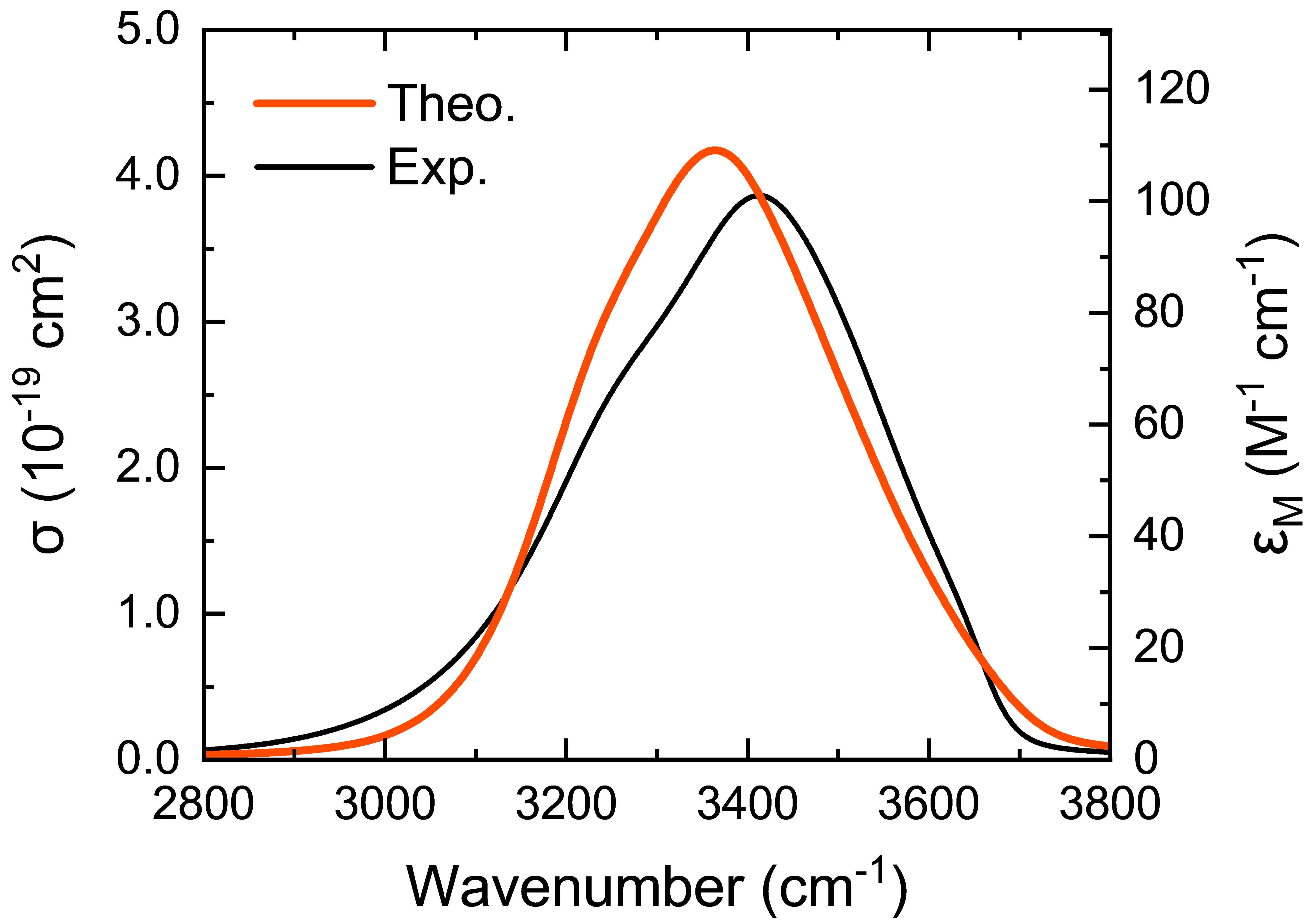}
    \caption{Theoretical and experimental values of absorption cross-section $\sigma$ (left vertical axis) and molar absorption coefficient $\epsilon_\text{M}$ (right vertical axis).}
    \label{fig:sigma}
\end{figure}

Now consider quantitative analysis of the spectra.
The theoretical peak height of $k$ is 0.328, in good agreement with the experimental value of 0.300 (Figure~\ref{fig:n-k}).
For $n$, the theoretical maximum and minimum values are 1.54 and 1.14, respectively, which also compare well with the experimental values (1.49 and 1.13).
The small deviation at the maximum corresponds to the deviation of the peak height of $k$ through the KK relation.
Other spectroscopic parameters, such as the absorption cross-section $\sigma$ and molar absorption coefficient ($\epsilon_\text{M}$), can be calculated from $k$ using the relations
\begin{align}
    \sigma            & = \frac{2\omega k}{c N}, \label{eq:sigma}                                             \\
    \epsilon_\text{M} & = \frac{2\omega k N_A}{c N \ln 10 } = \frac{N_A}{\ln 10} \sigma, \label{eq:epsilon_m}
\end{align}
where $c$ is the speed of light in a vacuum, $N$ is the number density of \ce{H2O} molecules, and $N_A$ is Avogadro's number.
The band strength ($\beta$) represents the total absorption strength of an absorption band, defined as an integral of $\sigma$ over the absorption band:
\begin{align}
    \beta = \int \dd{\omega} \sigma(\omega).
    \label{eq:A}
\end{align}
These values are used to quantify molecular abundance from IR spectra in fields such as astronomy and biology\cite{pontoppidanMappingIcesProtostellar2004,thiVLTISAAC35Mm2006,smithCospatialIceMapping2025,lorenz-fonfriaInfraredDifferenceSpectroscopy2020}.
Figure~\ref{fig:sigma} shows the theoretical values of $\sigma$, with corresponding values of $\epsilon_\text{M}$ on the right ordinate.
The applied theoretical density was obtained from the average number density in the MD trajectory.
Experimental values were calculated using data from Ref.~\citenum{maxIsotopeEffectsLiquid2009} and the experimental density of \SI{997}{\kilogram\per\meter\cubed} for water at \SI{300}{K}\cite{wagnerIAPWSFormulation19952002}.
Note that as $\sigma$ and $\epsilon_\text{M}$ are both proportional to $\omega k$, their line shapes are identical.
The accuracy of the calculated $\sigma$ (and $\epsilon_\text{M}$) appears similar to that of $k$, as expected from their proportionality and the relatively small contribution of the variation of $\omega$ to the spectra (see eqs~\eqref{eq:sigma} and \eqref{eq:epsilon_m}).
The ensuing calculated values of $\beta$ are \SI{1.53e-16}{cm} for theoretical results and \SI{1.51e-16}{cm} for experimental results, which are in excellent agreement.
This is likely because $\beta$, as an integrated quantity, is insensitive to errors in the predicted line shape.
Further verification with different materials and conditions is required to confirm the reliability of the method.

\begin{figure}
    \centering
    \includegraphics[]{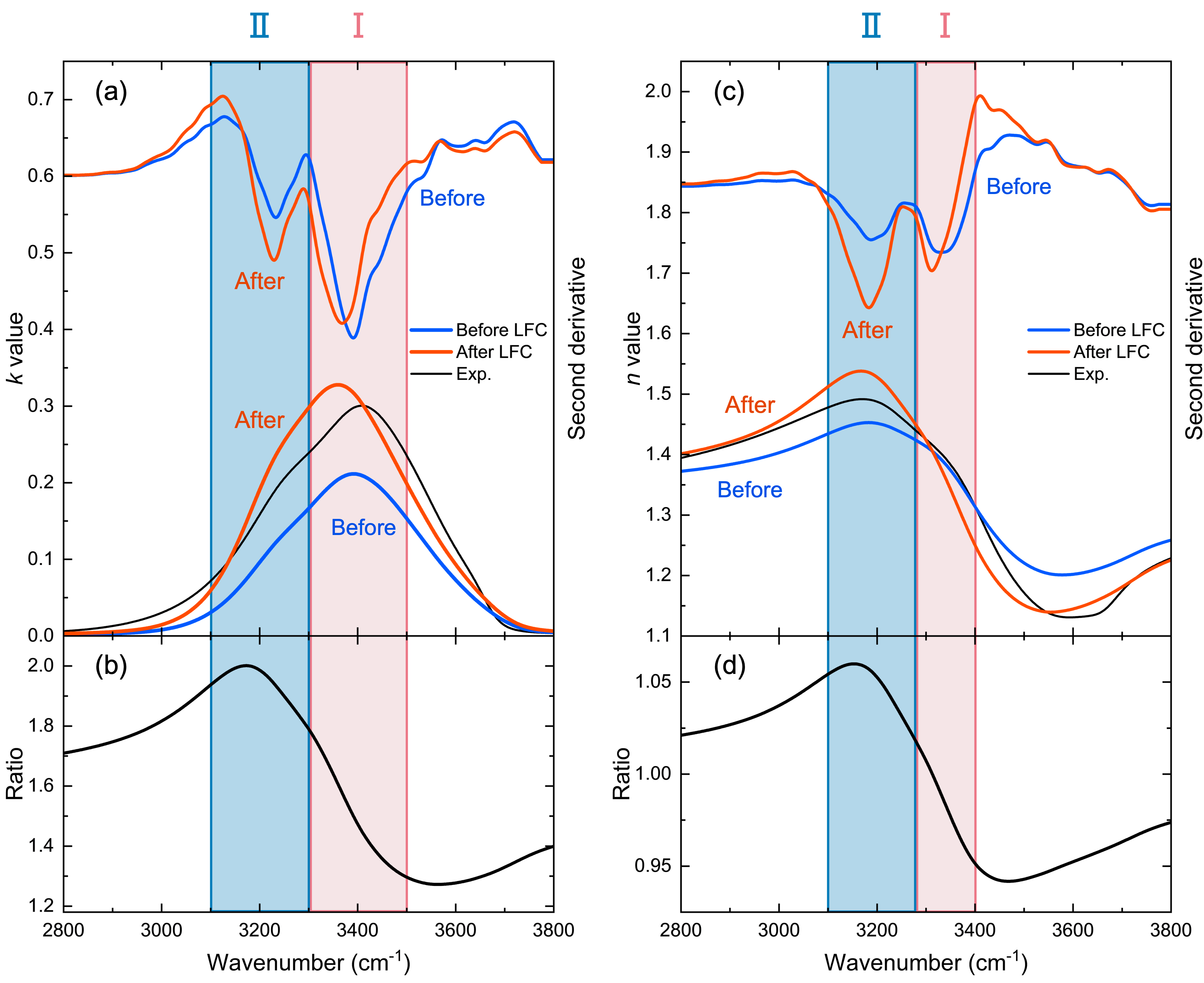}
    \caption{(a, c) Comparison of theoretical spectra of $k$ (a) and $n$ (c) before (blue) and after (red) the application of LFC\@.
        Top and middle spectra represent the second-derivative and original spectra, respectively.
        The second-derivative spectra of $k$ are calculated using normalized spectra of $k$ to have the same height.
        (b, d) Ratio of the corresponding theoretical spectra after LFC to that before LFC.}
    \label{fig:k-ba}
\end{figure}

To consider the effects of LFC, we compare spectra calculated with and without it.
Figure~\ref{fig:k-ba} shows theoretical spectra of $k$ and $n$, and their second derivatives, both before (blue) and after (red) application of LFC, alongside the experimental spectra (black).
The ratios of the two theoretical spectra (that after LFC divided by that before LFC) are shown in the lower panels.
The colored regions I and II are identical to those in Figure~\ref{fig:n-k-second}.

Figure~\ref{fig:k-ba}(a) shows the results for $k$.
Comparing features in the second-derivative spectra, the LFC red-shifts peak I by \SI{20}{\per\cm}, moving it away from the experimental position, whereas the position of peak II is largely unchanged.
LFC also alters the relative sizes of peaks I and II\@.
It diminishes peak I while enlarging peak II, consequently reducing the slope in the original spectrum at \qtyrange{3200}{3400}{\per\cm}.
These findings indicate that the LFC unevenly modifies the spectrum.
It also changes the peak height of $k$ from 0.212 (71\% of the experimental value) to 0.327 (109\% of the experimental value), an increase of 54\%.
Consider the $\beta$ values derived using eq~\eqref{eq:A}: the theoretical values before and after applying LFC are \SI{0.98e-16}{cm} (64\% of the experimental value) and \SI{1.53e-16}{cm} (101\% of the experimental value), respectively, an increase of 56\%.
Therefore, the introduction of LFC substantially improves the quantitative accuracy of $k$.

The above observations are clearly manifested in the ratio spectrum in Figure~\ref{fig:k-ba}(b).
The large slope in region I of the ratio spectrum corresponds to the red-shift in the original spectrum, whereas the near-zero average slope in region II of the ratio spectrum indicates little effect on the peak position.
The ratios are larger in region II than in region I, reflecting the selective enhancement of peak II\@.
These uneven contributions by LFC to the spectrum originate from the wavenumber-dependent enhancement of the local field, as represented in eq~\eqref{eq:loc-ext}.
In addition, the ratio spectrum is greater than 1 across the absorption band (1.6 on average), which corresponds to LFC overall increasing the intensity\@.
This is because with LFC, the absorption is evaluated according to $\vec{E}_\text{loc}$ instead of $\vec{E}_\text{ext}$, where the intensity of the former, $|\vec{E}_\text{loc}|$, is larger than that of the latter, $|\vec{E}_\text{ext}|$ (see SI for proof of this relation).

Figure~\ref{fig:k-ba}(c) shows the results for $n$.
As expected from the KK relation, the effects on $n$ of the LFC are similar to those on $k$.
In the second derivative, peak I red-shifts by \SI{10}{\per\cm}, moving away from the experimental position, whereas peak II remains unchanged.
However, peak II is more enhanced by the LFC\@ than is peak I.
These results show that LFC also modifies the spectrum of $n$, like that of $k$, in a non-uniform manner.
Applying LFC also increases the magnitude of $n$: the difference between the maximum and minimum values changes from 0.254 (70\% of the experimental value) to 0.404 (112\% of the experimental value), an increase of 59\%.

Similar to $k$, the above results are reflected in the relevant ratio spectrum (Figure~\ref{fig:k-ba}(d)).
The steep slope in region I of the ratio spectrum corresponds to the red-shift of peak I, and the small variation in region II reflects the small movement of peak II\@.
The larger value in region II compared with region I reflects the selective enhancement of peak II by LFC\@.
At the maximum and minimum of the original spectrum, the ratio values are 1.06 (>1) and 0.94 (<1), respectively.
Both of these factors act to increase the overall magnitude of the spectrum of $n$.
Considering the KK relation, the origin of both the selective modulation and the overall enhancement can be attributed to the same origin as the changes in $k$.

The results for both $k$ and $n$ demonstrate that LFC plays a major role in reproducing the quantitative spectra with the mixed quantum/classical approach.
Although the spectral intensity is also governed by the transition dipole moments, the error coming from the spectroscopic map to calculate them (eq~(1) in Ref.~\citenum{shiInterpretationIRRaman2012}) is relatively small.
As seen in Figure~1 of Ref.~\citenum{shiInterpretationIRRaman2012}, the fitting error of the map is minor compared to the large discrepancy of approximately 50\% in intensity between the experimental and theoretical spectra before applying the LFC (Figure~\ref{fig:k-ba}).
This stark difference, which is largely resolved by applying LFC, distinguishes the influence of local field effect on the absorption intensity from that of the transition dipole moments.

In conclusion, we have presented an extended version of Skinner's mixed quantum/classical approach and successfully reproduced the full complex refractive index ($k, n$), absorption cross-section ($\sigma$), molar absorption coefficient ($\epsilon_\text{M}$), and band strength ($\beta$) of liquid water at accuracies comparable to experiment.
The LFC was a crucial part of our methodology, substantially improving the predicted magnitudes of $k$ and $n$.
The theoretical band strength agreed particularly strongly with the experimental result, likely because this integrated quantity is insensitive to errors in the line shape.
A detailed evaluation considering the second derivative validated our extended approach as a tool for calculating both the macroscopic and microscopic optical properties of water.

The extended approach could easily be applied to other states of water, such as supercooled water, crystalline ice, and amorphous water, or to isotopologues of such states of water.
Using quantitative calculation of spectral intensity to re-evaluate the many theoretical studies that utilized Skinner's mixed quantum/classical approach \cite{schmidtAreWaterSimulation2007,auerIRRamanSpectra2008,liInfraredRamanLine2010,shiInterpretationIRRaman2012,tainterHydrogenBondingOHStretch2013,tainterStructureOHstretchSpectroscopy2014,niIRSFGVibrational2015,kananenkaFermiResonanceOHstretch2018,takayamaTheoreticalExperimentalODstretch2024} could deepen our understanding of the physical properties of water.
Amorphous water in particular has many structural uncertainties, and could benefit from our extended approach\cite{bartels-rauschIceStructuresPatterns2012}.
For example, vapor-deposited amorphous water formed at low temperature has been extensively studied in the field of astrochemistry because of its importance as a laboratory analog of interstellar icy dust grains.
Microscopic grains of ice play an important role in chemical evolution in space, serving as precursors of planetary material\cite{hamaSurfaceProcessesInterstellar2013,hamaStatisticalOrthotoparaRatio2016,nagasawaInfraredMultipleangleIncidence2022}.
As the typical thickness of vapor-deposited amorphous water is less than hundreds of nanometers, it is studied using IR spectroscopic approaches for thin films; e.g., normal-incidence transmission, reflection absorption, or multiple-angle incidence resolution spectroscopy\cite{nagasawaAbsoluteAbsorptionCross2021,nagasawaInfraredMultipleangleIncidence2022,hasegawaInfraredBandStrengths2024}.
A challenge facing these techniques is that the spectra are related to both $k$ and $n$, unlike the simple Beer--Lambert law for bulk measurements (eq~\eqref{eq:Lambert-Beer})\cite{hasegawaQuantitativeInfraredSpectroscopy2017,nagasawaAbsoluteAbsorptionCross2021,nagasawaInfraredMultipleangleIncidence2022,hasegawaInfraredBandStrengths2024,satoThreeStepGrowthVaporDeposited2025}.
Because our method explicitly provides values of both $k$ and $n$, it enables the direct calculation of spectra for thin-films of various phases of water, including vapor-deposited amorphous water.
This will be the subject of future work.

Our new results can be also applied to the direct investigation of the optical properties of sub-micron water clusters, such as interstellar icy dust grains or atmospheric aerosols.
Due to light scattering, their absorption spectra also become functions of both $k$ and $n$\cite{bohrenAbsorptionScatteringLight2008,hashmonayApplicationOpenPathFourier1999,medcraftWaterIceNanoparticles2013,tazakiLIGHTSCATTERINGFRACTAL2016,tazakiWatericeFeatureNearinfrared2021}.
Our method enables the simulation of the absorption spectrum of the entire cluster consisting of relatively large number of water molecules, thanks to the efficiency.
This capability is particularly relevant for interstellar icy dust grains (largely composed of amorphous water), as their molecular and morphological structure remains elusive despite the predominant role in determining the adsorption and reaction property of the their surface\cite{hamaSurfaceProcessesInterstellar2013}.
In fact, scattering features around the OH stretching and some other absorption bands have been captured by astronomical IR observations, including the recent high-quality spectra from the James Webb Space Telescope\cite{boogertObservationsIcyUniverse2015,mcclureIceAgeJWST2023,dartoisSpectroscopicSizingInterstellar2024,nobleDetectionElusiveDangling2024}, underscoring the significance of including scattering effects into the spectral simulation.
We expect that our extended method provides a powerful tool that will lead to new insights in the fields of astrochemistry and atmospheric chemistry.

Further improvements could be achieved by exploring the water model used in the MD simulation.
For example, Skinner and co-workers developed the explicit three-body (E3B) water model as a rigid water model based on the TIP4P model\cite{kumarWaterSimulationModel2008,tainterRobustThreebodyWater2011,tainterReparametrizedE3BExplicit2015}.
It explicitly considers three-body interactions, and
given its promising results\cite{pieniazekInterpretationWaterSurface2011,shiInterpretationIRRaman2012,tainterHydrogenBondingOHStretch2013,tainterStructureOHstretchSpectroscopy2014,niIRSFGVibrational2015,kananenkaFermiResonanceOHstretch2018}, replacing the TIP4P/2005 model with the E3B model could improve the performance of our calculations.
Intrinsic errors in the vibrational spectroscopic maps could also be reduced by, for example, increasing the number of parameters used to evaluate the environment around a given chromophore, which is currently only a single electric field along the OH bond\cite{baizVibrationalSpectroscopicMap2020,kitamuraTheoreticalAnalysisBetter2022}.
Given the additional consideration of spectral intensity, our approach might reveal improvements in performance from sophisticated maps that were not previously apparent when only the line shape was considered.

\section{Computational Details}
We theoretically calculated the vibrational spectra of water according to the mixed quantum/classical approach developed by Skinner's group\cite{auerDynamicalEffectsLine2007,auerIRRamanSpectra2008,gruenbaumRobustnessFrequencyTransition2013,kananenkaFermiResonanceOHstretch2018}, with modifications to obtain quantitative and complex spectra. Prior to calculating the spectra, we performed classical MD simulations with the GROMACS 2024.2 package\cite{abrahamGROMACSHighPerformance2015}.
The system consisted of 512 molecules of the TIP4P/2005 water model.
It was maintained at a constant temperature of \SI{298}{K} using the Nosé--Hoover thermostat\cite{noseMolecularDynamicsMethod1984,hooverCanonicalDynamicsEquilibrium1985} and at a pressure of \SI{1}{\bar} using the stochastic cell rescaling barostat\cite{bernettiPressureControlUsing2020}.  The equations of motion were integrated using the leap-frog algorithm with a \SI{2}{fs} time step.  A real-space cutoff length of \SI{1.0}{nm} was applied to long-range Coulomb interactions, which were calculated using the particle mesh Ewald method\cite{essmannSmoothParticleMesh1995,dardenParticleMeshEwald1993}.
Lennard--Jones interactions were truncated at \SI{1.0}{nm}, and the standard long-range corrections were included in the calculations of pressure and the potential energy. Atomic coordinates were obtained every \SI{2}{fs} for \SI{76}{ps} after \SI{100}{ps} of equilibration.

Following the MD simulation, we calculated the IR spectra using eqs~\eqref{eq:eps-pre-loc} and \eqref{eq:fin}.  We chose \SI{76}{fs} for the averaging time $T$ to best reproduce motional narrowing\cite{auerDynamicalEffectsLine2007}.
Although Skinner's group used a $T_1$ of \SI{260}{fs} to incorporate lifetime broadening effects\cite{lockTemperatureDependenceVibrational2002,auerIRRamanSpectra2008}, we set the full width at half maximum of the Lorentzian function in eq~\eqref{eq:lorentz} to \SI{40}{\per\cm}, a value which yields better agreement with the experimental line shape\cite{takayamaTheoreticalExperimentalODstretch2024}.
The value of $V$ used in eq~\eqref{eq:eps-pre-loc} was obtained by averaging the volume of the simulation box during the production run.
The background dielectric constant ($\epsilon_b$) was set to a value at which $\epsilon_\text{cor}$ becomes the square of the background refractive index ($n_\text{el}$), which is the component of the refractive index originating solely from electronic polarization.
The value of $n_\text{el}$ was calculated as follows:\cite{bertieRefractiveIndexColorless1995}
\begin{align}
    n_\text{el} (\tilde{\nu}) = 1.32663 + \num{2.439e-11} \tn^2 + \num{3.74e-21} \tn^4,
\end{align}
where $\tn$ is the wavenumber in \si{\per\cm}.
Our present calculations have $\tn$ set to $\SI{3800}{\per\cm}$, which yielded $\epsilon_b = 1.607$.
Values used to calculate the Hamiltonian and transition dipole moments were obtained from the vibrational spectroscopic maps developed for the TIP4P model\cite{jorgensenComparisonSimplePotential1983tip4p} by Skinner and co-workers \cite{gruenbaumRobustnessFrequencyTransition2013,kananenkaFermiResonanceOHstretch2018}.
The transferability of these maps for use with the TIP4P/2005 model has been previously established by Takayama et al.\cite{takayamaTransferabilityVibrationalSpectroscopic2023}
A comparison of spectra calculated with TIP4P, TIP4P/2005, and TIP4P/Ice\cite{abascalPotentialModelStudy2005tip4pIce} under identical conditions to the present study are shown in Figure~S3.
Although the TIP4P model yields slightly better agreement with the experiment, we have chosen TIP4P/2005 due to its versatile performance across a wide range of temperatures and phases\cite{vegaSimulatingWaterRigid2011}.
This choice is motivated by our intent to apply the present method to future studies of various states and temperatures as mentioned in the text.

\section*{Notes}
The authors declare no competing financial interest.

\section*{Acknowledgments}
We thank Dr. Atsuki Ishibashi for useful discussions.
This work was supported by grants from the Japan Society for the Promotion of Science KAKENHI (24H00264) and the JST FOREST Program (JPMJFR231J).

\section{TOC Graphics}
\begin{figure}[H]
    \centering
    \includegraphics[width=0.6\linewidth]{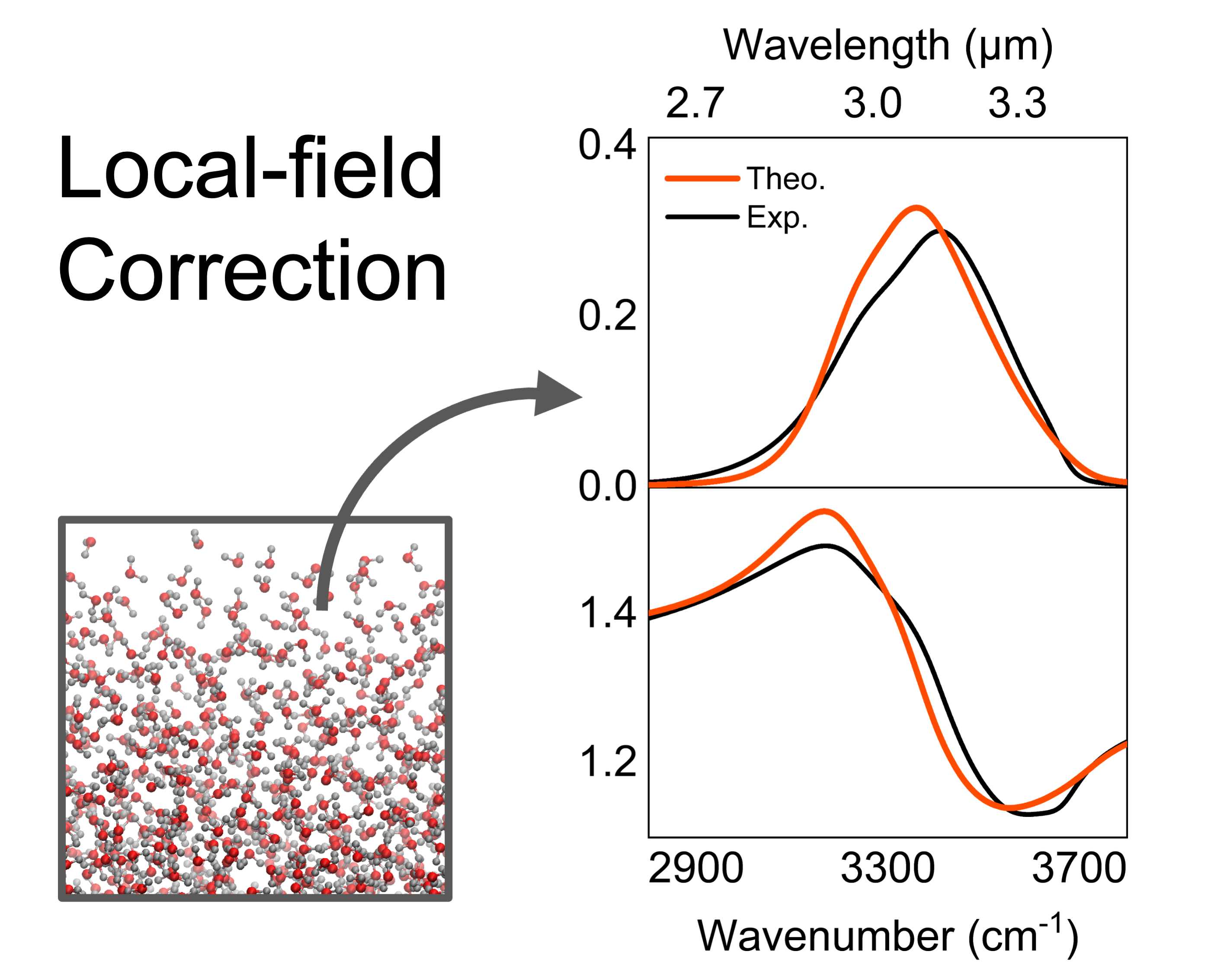}
\end{figure}

\section*{Supporting Information}
The following files are available free of charge.

A detailed theorization of the relation between local electric field ($\vec{E}_\text{loc}$) and external electric field ($\vec{E}_\text{ext}$) in terms of their intensities; experimental spectra of normal-incidence transmission and reflection-absorption spectroscopy, complex refractive index ($n+ik$), and calculated transverse and longitudinal optic energy-loss functions ($f_\text{TO}$, $f_\text{LO}$) of amorphous water at 10 K (Figure S1); theoretical and experimental values for complex dielectric constant ($\epsilon_1 + i\epsilon_2$) of liquid water (Figure S2); theoretical values for $n$ and $k$ of liquid water calculated using TIP4P, TIP4P/2005, and TIP4P/Ice water models (Figure S3).

\section*{Acknowledgments}

\bibliography{paper}

\end{document}